\documentclass[preprint,12pt]{elsarticle}
\usepackage{dcolumn}%
\usepackage{bm}%
\usepackage{graphicx}
\usepackage{amsfonts}
\usepackage{amsmath}
\usepackage{amssymb}
\usepackage{hyperref}
\usepackage{indentfirst}
\usepackage[nodots]{numcompress}
\usepackage[usenames,dvipsnames]{pstricks}
\usepackage{epsfig}
\usepackage{pst-grad} 
\usepackage{pst-plot} 
\journal{Physics Letters A}

\begin{document}
\begin{frontmatter}

\title{Stability of skyrmions  on curved surfaces in the presence of a magnetic field}
\author[1,2]{V.L. Carvalho-Santos}
\address[1]{Instituto Federal de Educa\c c\~ao, Ci\^encia e Tecnologia Baiano - Campus 
Senhor do Bonfim, \\Km 04 Estrada da Igara, 48970-000 Senhor do Bonfim, Bahia, Brazil}
\address[2]{Departamento de F\'isica, Universidad de Santiago de Chile and CEDENNA,\\ Avda. Ecuador 3493, Santiago, Chile}
\author[2]{R.G. Elias}
\author[2]{D. Altbir}
\author[3]{J.M. Fonseca}
\address[3]{Universidade Federal de Vi\c cosa, Departamento de F\'isica, 
Avenida Peter Henry Rolfs s/n, 36570-000, Vi\c cosa, MG, Brasil}

\begin{abstract}

We study the stability and energetics associated to skyrmions appearing as excitations on curved surfaces. Using a continuum model we show that the presence of cylindrically radial and azimuthal fields destabilizes the skyrmions that appear in the absence of an external field.  Weak fields generate fractional skyrmions while strong magnetic fields yield stable $2\pi$-skyrmions, which have their widths diminished by the magnetic field strength. Under azimuthal fields vortex appear as stable states on the curved surface.

\end{abstract}

\end{frontmatter}

\section{Introduction}
During the last decade a strong interest has focused on the properties of topological structures on curved surfaces based on the fact that the surface's shape determines the physical properties of several systems. As examples, one can cite the ordering of nematic liquid crystals on a curved substrate \cite{Napoli-PRL-2012} and the local density of states of a conical graphene sheet \cite{jakson-graphene}. In the same hand, curvature is important to describe the physical properties of magnetic nanostructures. In a recent work by Carvalho-Santos {\it et al}, it was shown that toroidal nanorings can support a vortex-like magnetization for smaller sizes than cylindrical nanorings \cite{Vagson-JAP-2010}. 

Regarding recent experiments, curved nanostructures have been prepared. As examples we can mention the production of permalloy caps on non-magnetic spheres \cite{Streubel-APL-2012} and permalloy cylindrically curved magnetic segments with different radii of curvature on non-magnetic rolled-up membranes \cite{Streubel-Nanoletters-2012}. Moreover, the synthesis of periodically modulated nanowires, which can be controlled by external
parameters that allow control of the geometry of pores has been reported in several works (See \cite{knielsch-works} and references therein). 

Skyrmions were first described in nuclear physics, when Skyrme showed that topologically stable field configurations for interacting pions can occur as particle-like solutions \cite{skyrme-original}. However, these topological structures do not appear only in particle Physics. In the last two decades, skyrmions have been studied in several condensed matter systems, e.g., nematic liquid crystals \cite{Pu-Op-2013}, Bose-Einstein condensates \cite{Khawaja-Nature-2001}, and magnetic systems (see Ref. \cite{Iwasaki-Nat-2013,Nature-2013} and references therein). In particular, regarding magnetic systems, skyrmions are chiral spin structures with a whirling configuration which appear like a magnetization groundstate due to the competition among Exchange, anisotropy, Zeeman and Dzialoshinsky–Moriya interactions \cite{Nature-2013}. Otherwise, skyrmions may appear as excited state of the continuum Heisenberg model (HM), which consists in the non-linear $\sigma$-model if the constraint $m^2=1$ is considered to the spin space \cite{Rajaraman}. These spin collective modes have topological stability since its structure cannot be continuously deformed to a ferromagnetic or other magnetic state. When present in curved systems, the energy, stability and width associated to skyrmion-like excitations depend on the surface's curvature \cite{cylinder,sphere,pseudosphere,torusmeu,Carvalho-Santos-PLA,Dandoloff-JPA-2011}.

Related to curved systems, it has been previously shown that in the presence of an axial magnetic field  $2\pi$-skyrmions appear in simply and non-simply connected surfaces \cite{Vagson-Dandoloff-PLA,Saxena-PRB-58,Boas-PLA-2014}. Based on these ideas, in this paper we extend these results by studying the excitations appearing on magnetic surface with rotational symmetry  under  a radial or an azimuthal magnetic field. Our results show that a weak magnetic field can breakout the skyrmion stability, by shifting the energy minima in the absence of fields. On the other hand, a strong magnetic field coupled to the curvature of the surface can lead to the appearance of $2\pi$-skyrmions. The magnitude and direction of the magnetic field play an important role on properties such as the phase, the width and the energy of the skyrmions.

This work is divided as follows: in section \ref{sec2} we develop the Heisenberg Model on cylindrically symmetric surfaces in the absence and presence of an external magnetic field. In Sec. \ref{MagFieldSection}, we present the derived equations coming from the interaction with cylindrically radial (RMF) and azimuthal magnetic fields (AMF). Sec. \ref{Results} presents an analytical study of the appearance and stability of 2$\pi$-skyrmions due the interaction with AMF an RMF. Finally, in section \ref{conclusions} we present the conclusions and prospects of this work.

\section{Continuum Heisenberg Model on cylindrically symmetric surfaces}\label{sec2}
A spin system lying on a surface embedded in a 3D-space can be described by a Heisenberg Model in a continuum approximation of spatial and spin variables, valid at sufficiently large wavelength and low temperature. Then, and assuming only exchange terms, we can describe our spin system by the following Hamiltonian \cite{cylinder}
\begin{equation}
\label{heiscont} 
H=J\iint g^{\mu\nu}{\partial_\mu m^{\alpha}}{\partial_\nu m_{\alpha}} dA,
\end{equation} 
where $\partial_{\mu,\nu}\equiv\partial/\partial\eta^{\mu,\nu}$. The surface is described by the curvilinear coordinates $\eta_{1}$ and $\eta_{2}$, $dA=\sqrt{|det[g_{\mu\nu}]}d\eta_{1}d\eta_{2}$ is the surface element, $g^{\mu\nu}$ is the surface contravariant metric. Repeated indices must be summed, with $\mu$ and $\nu$ varying from 1 to 2 and $\alpha$ varying from 1 to 3. $J$ denotes the coupling between neighboring spins, and according to $J<0$ or $J>0$, the Hamiltonian describes a ferro or antiferromagnetic system. The classical spin vector field is given by $\vec{m}=(\sin\Theta\cos\Phi,\sin\Theta\sin\Phi,\cos\Theta)$, so that $\Theta=\Theta(\eta_{1},\eta_{2})$ and $\Phi=\Phi(\eta_{1},\eta_{2})$. 

\begin{figure}
\includegraphics[scale=0.27]{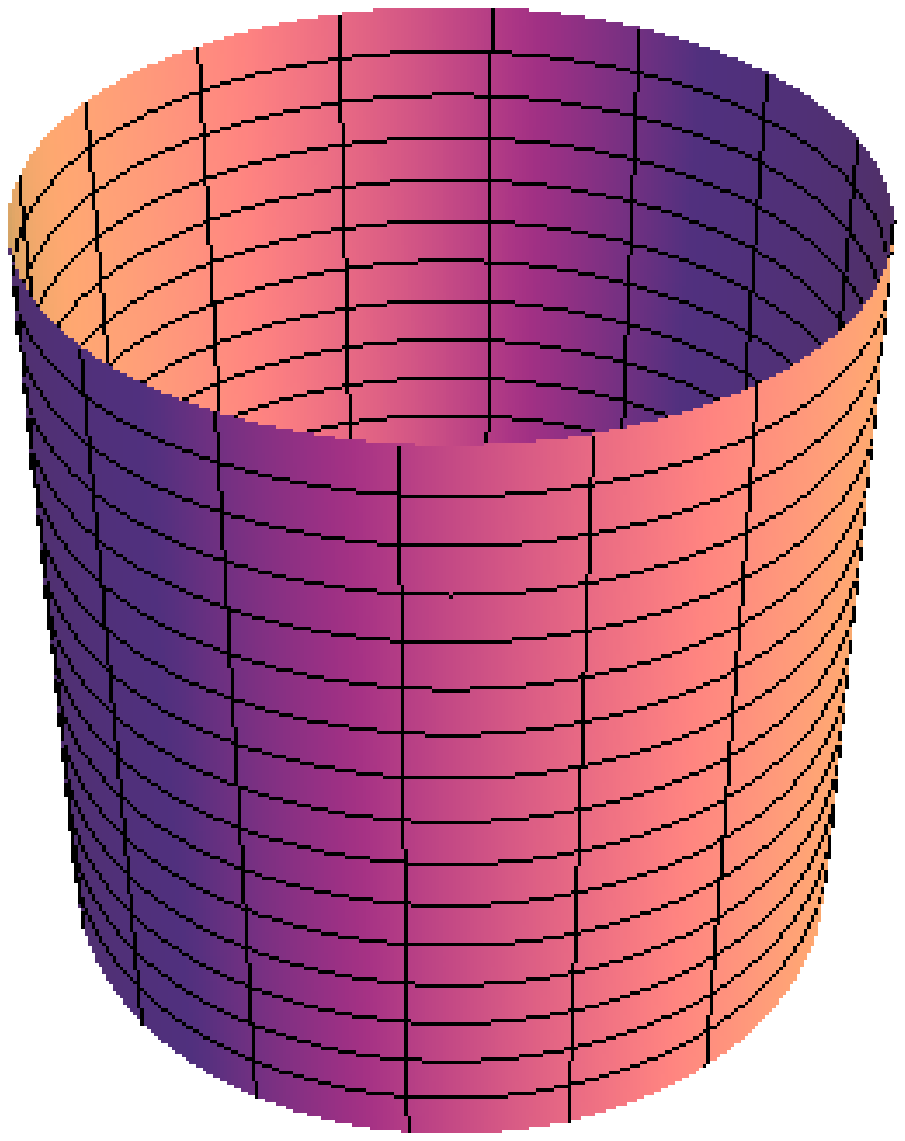}\hspace{1cm}\includegraphics[scale=0.31]{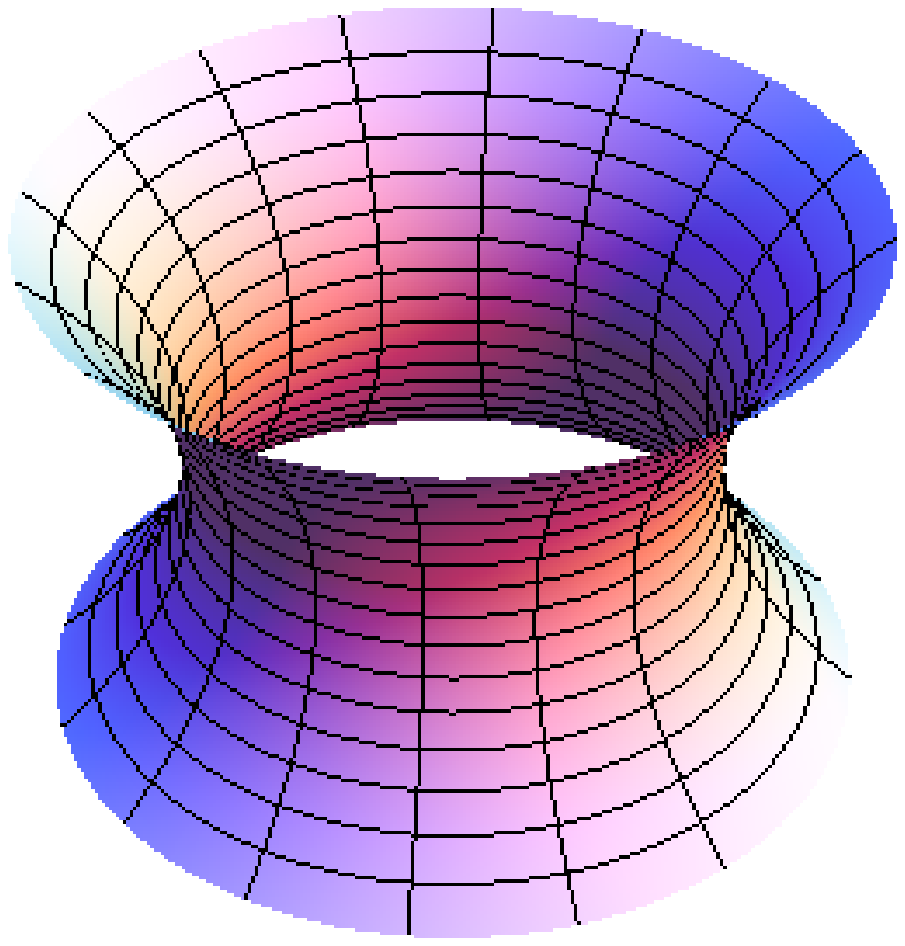}\hspace{1cm}\includegraphics[scale=0.4]{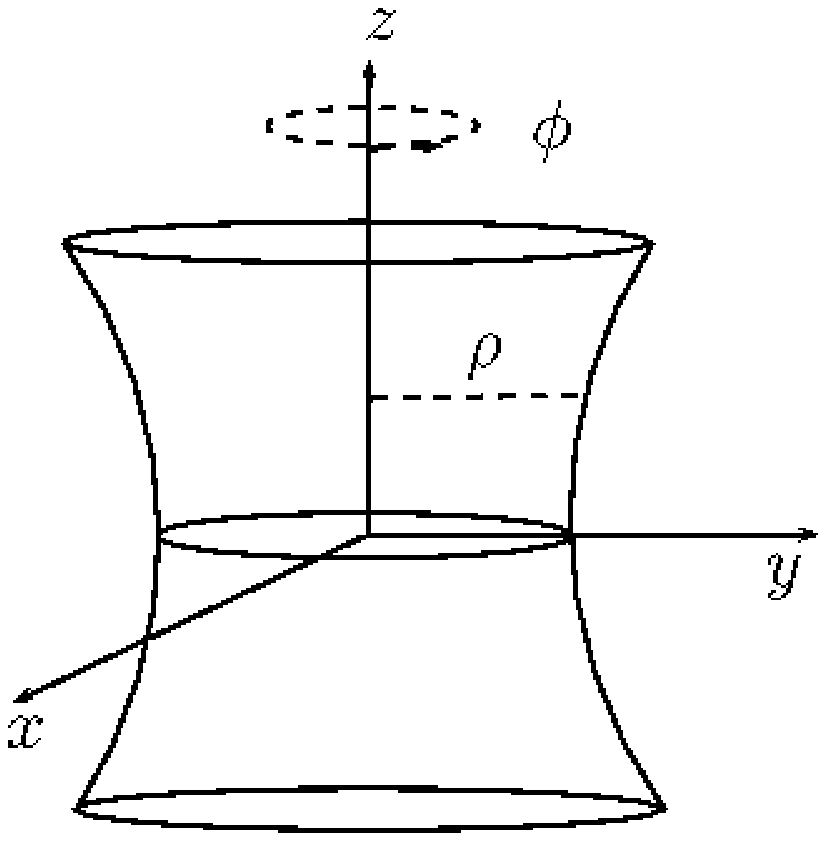}\caption{Examples of surface with cylindrical symmetry, a cylinder and a catenoid. At right, we show the representation of the cylindrical-like coordinate system of a surface with rotational symmetry. $\phi$ is the azimuthal angle and $\rho$ is the radius of the surface at the heigth $z$.}\label{Coordinate}
\end{figure}

Our main interest is to study static topological excitations appearing when we consider a rotationally symmetric surface in the presence of external fields. In cylindrical coordinate system, an arbitrary surface with rotational symmetry can be parametrized by $\mathbf{r}=(\rho,\phi,z(\rho))$, where $\rho$ is the radius of the surface at height $z$, and $\phi$ accounts for the azimuthal angle (See Fig. \ref{Coordinate}). Thus, the Hamiltonian (\ref{heiscont}) can be rewritten as
\begin{eqnarray}\label{HamGen}
H=J\iint\Big\{\sqrt{\frac{g_{\phi\phi}}{g_{\rho\rho}}}\left[(\partial_{\rho}
\Theta)^2+\sin^2\Theta(\partial_{\rho}\Phi)^2\right]\nonumber\\
+\sqrt{\frac{g_{\rho\rho}}{g_{\phi\phi}}}\left[(\partial_{\phi}
\Theta)^2+\sin^{2}\Theta(\partial_{\phi}\Phi)^2\right]\Big\}d\rho d\phi.
\end{eqnarray}

The Euler-Lagrange equations (ELE) derived from (\ref{HamGen}) are
\begin{eqnarray}\label{GenForm}
2(\partial^2_{\zeta}\Theta+\partial^2_{\phi}\Theta)=\sin2\Theta\left[\left(\partial_{\zeta}
\Phi\right)^2+\left(\partial_{\phi}\Phi\right)^2\right]
\end{eqnarray} 
and
\begin{eqnarray}\label{phieq}
\sin^2\Theta\left(\partial^{2}_{\zeta}\Phi+\partial^{2}_{\phi}\Phi\right)+\sin2\Theta\left(
\partial_{\zeta}\Theta\partial_\zeta\Phi+\partial_{\phi}\Theta\partial_\phi\Phi\right)=0 \, .
\end{eqnarray}

\noindent
In these expressions $d\zeta=\sqrt{{g_{\rho\rho}}/{g_{\phi\phi}}}d\rho$ is a length scale depending on the geometric parameters of the underlying manifold and we have used the fact that $\partial_{\phi}(\sqrt{g_{\phi\phi}/g_{\rho\rho}})=0$, since $g_{\phi\phi}=\rho^2$ and $g_{\rho\rho}=1+(dz/d\rho)^2$ have no dependence on $\phi$.

The set of equations describing the spin vector field present a length scale parameter that depends on the considered geometry, leading to shape induced changes in the energy and the stability of the excitations of our spin system. When a spin vector field  with cylindrical symmetry is considered, that is, $\Phi(\rho,\phi)\equiv\Phi(\phi)$ and $\Theta(\rho,\phi)\equiv\Theta(\rho)$, the above set of equations is simplified and the sine-Gordon system is obtained. Its solution is a topological $\pi$-skyrmion, represented by the spin profile given by $\Phi=\phi+\phi_0$ and
\begin{eqnarray}\label{piskyrmion}
\Theta_{\text{iso}}=2\arctan(\text{e}^{\zeta/\zeta_0})\,,
\end{eqnarray}
where $\phi_0$ and $\zeta_0$ are phases depending on the boundary conditions and that does not account to the energy calculations. If a skyrmion profile is given by a function $f(\lambda\zeta)$, its width is given by $\lambda^{-1}$, which generally appears in front of the $\sin2\Theta$ term of Eq. ({\ref{GenForm}). However, due to the chosen parametrization of the surface, the skyrmion width is rescaled to unity. Nevertheless, the curvature dependence of the skyrmion width can be evidenced by calculating $\zeta=\int{\sqrt{{g_{\rho\rho}}/{g_{\phi\phi}}}d\rho}$.

Once the minima associated to Eq. (\ref{HamGen}) are given by $\Theta=n\pi$, with $n$ integer, the $\pi$-skyrmion appears as a continuous transition connecting the two neighboring minima $0$ and $\pi$. This transition consists in a topological excitation, mapping the spin sphere on the underlying manifold, so belonging to the first class of the second homotopy group. Despite the ferromagnetic state is less energetic than the skyrmionic one, there is a very high energy barrier separating these states and the skyrmion can not be deformed in a ferromagnetic-like state by a continuous variation of the order parameter, characterizing it as a topological excitation. In the case of finite surfaces, Eq. (\ref{piskyrmion}) does not represent an integer $\pi$-skyrmion, since the spin sphere mapping is partially done and fractional skyrmions are obtained (See Refs. \cite{cylinder,Carvalho-Santos-PLA} for more details). The spin pattern of a $\pi$-skyrmion with phase $\phi_0=\pi/2$ is shown in Fig. \ref{SpinPattern}.(b). 
 
\begin{figure}
\includegraphics[scale=0.315]{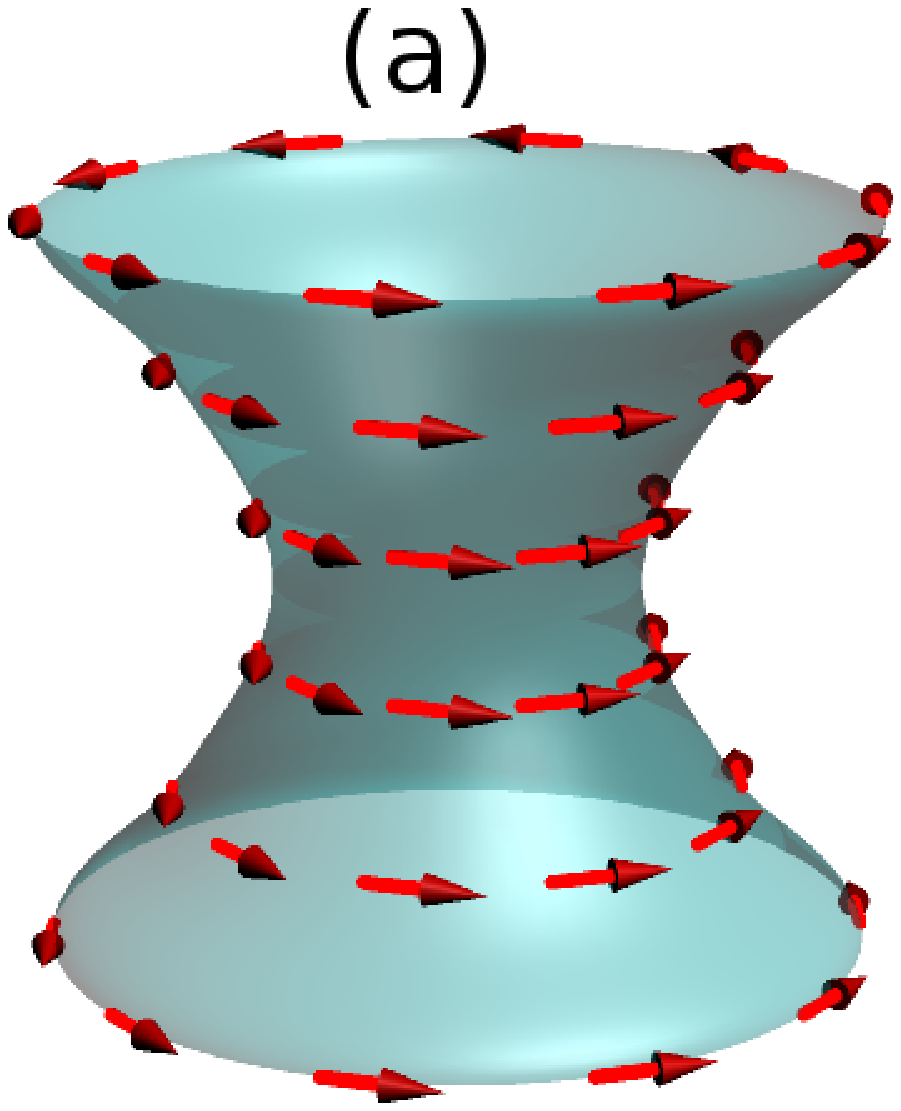}\hspace{0.5cm}\includegraphics[scale=0.343]{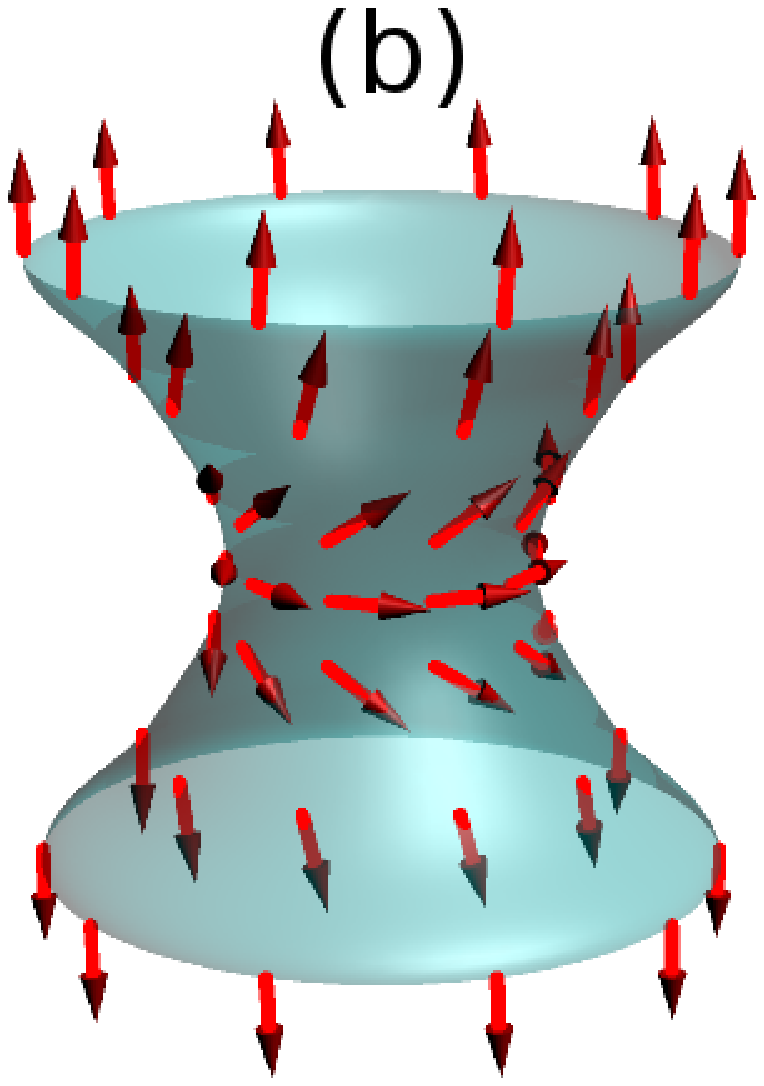}\hspace{0.5cm}\includegraphics[scale=0.315]{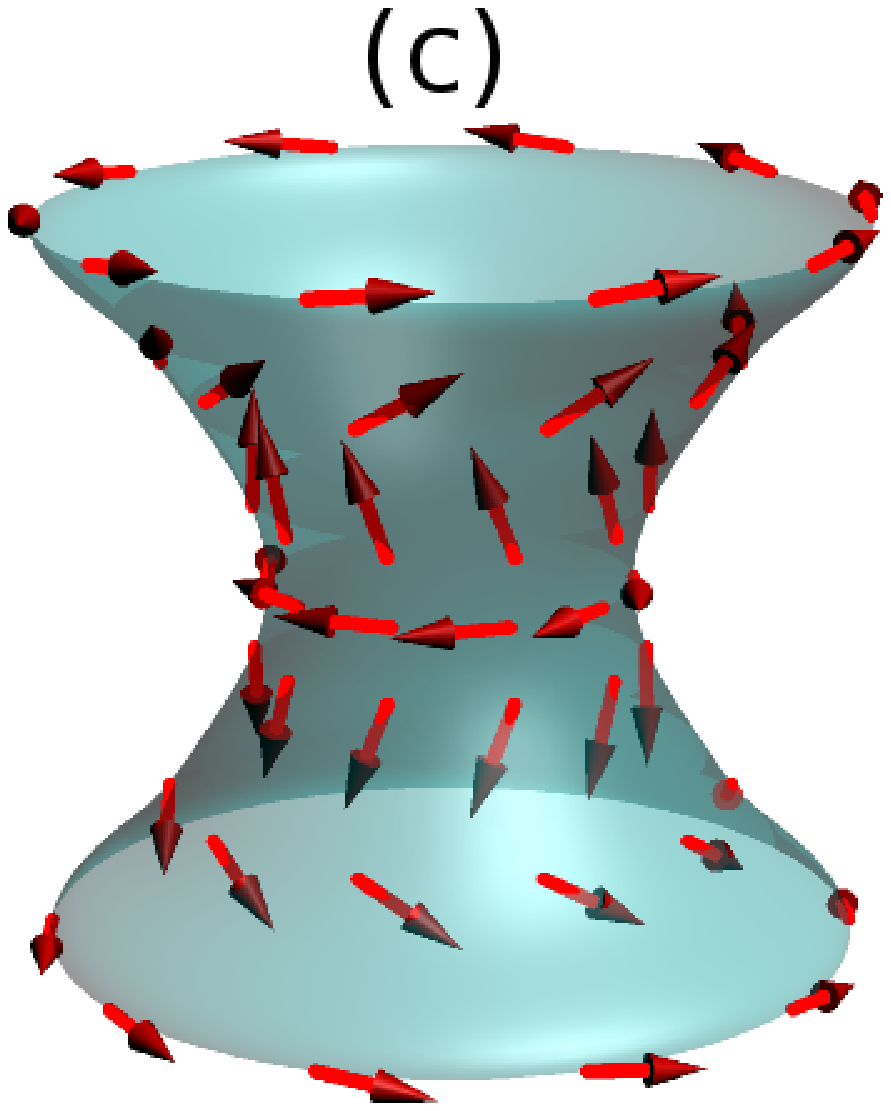}\hspace{0.5cm}\includegraphics[scale=0.315]{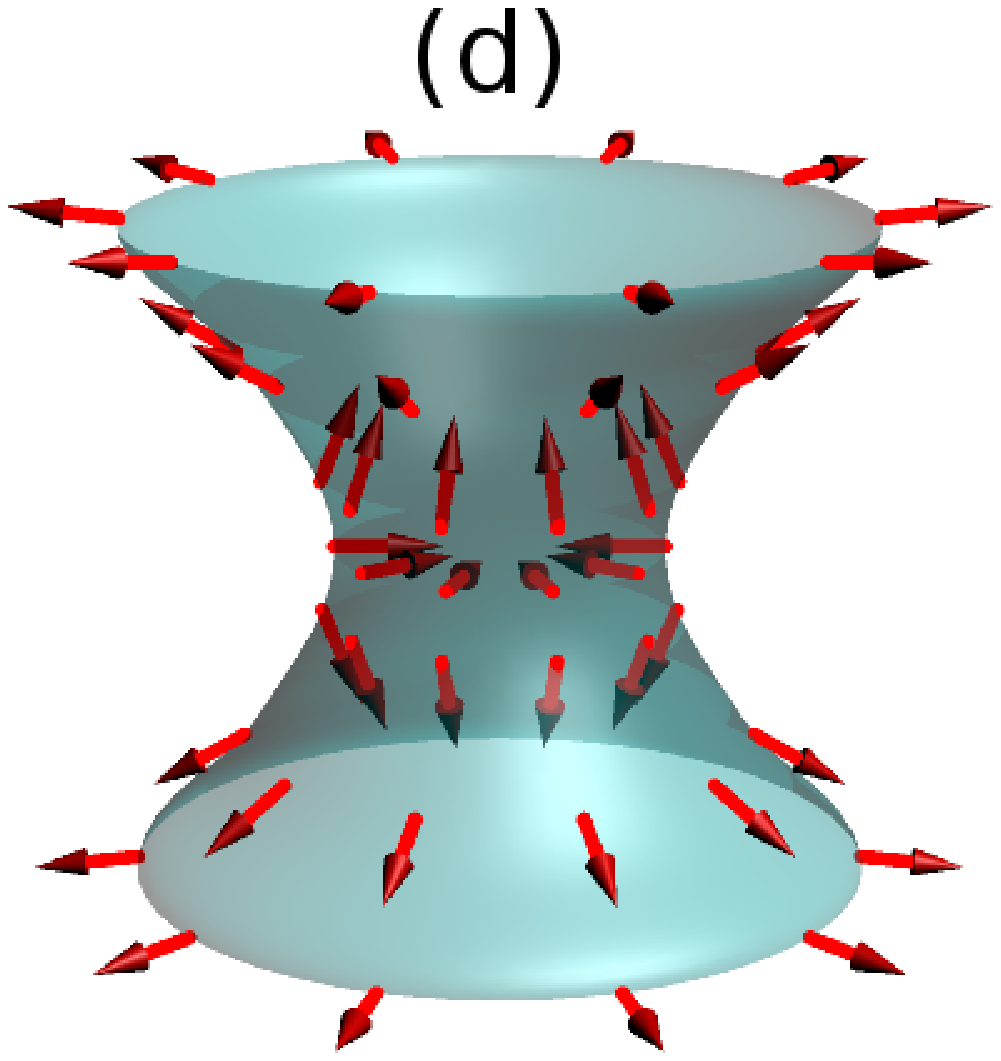}\caption{Spin patterns appearing on the hyperboloid, which is a curved surface with rotational symmetry. Figure (a) represents a vortex configuration, while figure (b) shows the spin pattern of a $\pi$-skyrmion with phase $\phi_0=\pi/2$. Figures (c) and (d) show $2\pi$-skymions pattern generated by the interaction with azimuthal (c) and radial (d) magnetic fields.}\label{SpinPattern}
\end{figure}

Another class of solution to Eqs. (\ref{GenForm}) and (\ref{phieq}) are given by $\Theta=\pi/2$ and $\Phi=\phi+\phi_0$. These solutions represent a vortex-like excitation, whose profile can be viewed in Fig. \ref{SpinPattern}.(a). The energy of the vortex can be obtained from Eq. \ref{HamGen},  giving $E_{v}=2\pi J\zeta$. This expression evidence the relation between the energy and the length scale of the surface.

\section{Interaction with external fields}\label{MagFieldSection}
When we include an external magnetic field, a new class of solutions  can be obtained. This interaction can be modeled by adding into the Hamiltonian (\ref{HamGen}) the term
\begin{equation}\label{HamMagInt}
H_{_\text{int}}=-g\mu\iint\mathbf{m}\cdot\mathbf{B}\,dA,
\end{equation}
where $\mathbf{B}$ is the applied magnetic field, $\mu$ is the magnetic moment, and $g$ is the Land\'e factor of the electrons in the magnetic materials. Spin systems described by a Heisenberg model in the presence of a constant axial external field have been previously studied on a circular cylindrical surface. In this case the ELE transform into a homogeneous double sine-Gordon equation (DSGE)\cite{Saxena-PRB-58}. The solutions to the homogeneous DSGE are $2\pi$-skyrmions,  that change  their widths  due to a second length scale introduced into the system by the magnetic field. If a constant axial magnetic field is interacting with a spin system on an arbitrary geometry, the non-homogeneous DSGE appears and the solutions can be obtained only numerically \cite{Dandoloff-JPA-2011}. However, if the strength of the axial magnetic field is tuned with the curvature of the substrate, the homogeneous DSGE is recovered \cite{Vagson-Dandoloff-PLA} and solutions are obtained analytically. 

In order to extend the above cited results, we will consider two distributions of magnetic fields: (a) a radial magnetic field (RMF), $\mathbf{B}=B(\rho)\hat{\rho}$, where $\hat{\rho}$ is the unitary vector in cylindrical coordinates; and (b) a Azimuthal magnetic field, AMF, i.e., $\mathbf{B}=B(\rho)\hat{\phi}$. 

\subsection{Cylindrically Radial Magnetic Field} 

From Eq.\ref{HamMagInt}, a magnetic system in the presence of a  RMF can be described by 
\begin{eqnarray}\label{HamIntRadial}
H_{\text{int}-\hat{\rho}}=-\iint\sqrt{\frac{g_{\rho\rho}}{g^{\phi\phi}}}B'(\rho)\sin\Theta\cos(\Phi-\phi)d\rho d\phi\,,
\end{eqnarray}
where $B'(\rho)=g\mu B(\rho)$. The derived ELE are
\begin{eqnarray}\label{ThetaRadEqMagExt}
\partial_{\zeta}^2\Theta=\frac{\sin2\Theta}{2}\left(\partial_\phi\Phi\right)^2-\frac{\rho^2}{2}
{B'(\rho)}\cos\Theta\cos(\Phi-\phi)
\end{eqnarray}
and
\begin{eqnarray}\label{PhiRadEqMagExt}
\sin\Theta\partial_{\phi}^2\Phi=-\rho^2 B'(\rho)\sin(\Phi-\phi)\,.
\end{eqnarray}

We start by considering $\Theta=\pi/2$ as a solution of Eq. (\ref{ThetaRadEqMagExt}). In this case, the solution of Eq. (\ref{PhiRadEqMagExt}) is given by $\Phi=\phi+n\pi$, with  $n$ integer. From Eq. (\ref{HamIntRadial}) it can be seen that in order to minimize the energy, $n$ has to be even, and therefore the spin vector field aligns parallel to the magnetic field. Thus, the radial magnetic field induces a constraint into the system, removing the freedom of the phase's choice that appears when no magnetic field is acting into the system. In order to study excitations of the order parameter $\Theta$, numerical solutions must be obtained to Eqs. (\ref{ThetaRadEqMagExt}) and (\ref{PhiRadEqMagExt}). However, since $\zeta$ depends on the geometry, numerical solutions must be performed separately to each surface (See for example, Ref. \cite{Boas-PLA-2014}). On the other hand, analytical solutions to Eq. (\ref{ThetaRadEqMagExt}) can be obtained for magnetic  fields of the form $B'_{0}/\rho^2$. In this case Eq. (\ref{ThetaRadEqMagExt}) can be rewritten as
\begin{eqnarray}\label{ThetaRadSimp}
\partial_{\zeta}^2\Theta=\frac{\sin2\Theta}{2}-\frac{1}{2}B'_{0}\cos\Theta\,. 
\end{eqnarray}

Despite the difficulty to generate a radial field varying with $1/\rho^2$, recent theoretical and experimental propositions can give some insights in this direction. One of them consists in using magnetic monopoles, which generate a radial magnetic field proportional to $1/r^2$, where $r$ is the distance from the monopole to the point at which the magnetic field is being measured. Magnetic monopoles are predict to appear in a pyrochlore lattice, which is an example of a ferromagnetic material with a degenerate ground state that is disordered at low temperatures. The magnetic monopoles are emergent particles appearing like an excited spin state, resulting of collective modes into these systems, also known as spin ices \cite{Castelnovo-Nature-2008}. However, despite the spin ice provide a system to study magnetic monopoles, it is not possible to separate the monopole from the material. On the other hand, in a recent experiment, using a nanoscopic ferromagnetic needle of Nickel, B\'ech\'e \textit{et al} measured the Aharanov-Bohm phase shift caused by the magnetic potential around the needle. Their results show that the needle tip behaves as a magnetic monopole, whose polarity can be chosen depending on the magnetization direction  \cite{Beche-Nature-2014}. This experiment follows the Dirac's proposition to generate a magnetic monopole by connecting two oppositely charged magnetic monopoles with an infinitesimal string of flux. The approximation to a magnetic monopole is made by pulling the monopoles as far as possible and providing means of localizing the return flux connecting both monopoles.

Equation (\ref{ThetaRadSimp}) will be denoted as the homogeneous double sine-cosine-Gordon equation (DSCGE) because it contains a $\cos\Theta$, while the DSGE presents a $\sin\Theta$ in the second term of the right side.

\subsection{Azimuthal magnetic field}

When an AMF is considered, the interaction with the  spin vector field with cylindrical symmetry can be written as
\begin{eqnarray}\label{HamMagExtInt}
H_{\text{int}-\hat{\phi}}=-\iint\sqrt{\frac{g_{\rho\rho}}{g^{\phi\phi}}}B'(\rho)\sin\Theta\sin(\Phi-\phi)d\rho d\phi\,.
\end{eqnarray}
In this case, the derived ELE are evaluated and give
\begin{eqnarray}\label{ThetaEqMagExt}
\partial_{\zeta}^2\Theta=\frac{\sin2\Theta}{2}\left(\partial_\phi\Phi\right)^2-\frac{\rho^2}{2}
{B'(\rho)}\cos\Theta\sin(\Phi-\phi)
\end{eqnarray}
and
\begin{eqnarray}\label{PhiEqMagExt}
\sin\Theta\partial_{\phi}^2\Phi=\rho^2 B'(\rho)\cos(\Phi-\phi)\,.
\end{eqnarray}

Note that Eq. (\ref{PhiRadEqMagExt}) can be obtained from Eq. (\ref{PhiEqMagExt}) by doing the transformation $\phi\rightarrow\phi+\pi/2$. Then, the simplest solution to Eq.  (\ref{PhiEqMagExt}) is given by $\Phi=\phi+(2n+1)\pi/2\,,$. Therefore, as well as in the radial field case, the presence of an AMF removes the freedom of the phase angle and the spins are forced to align parallel to the azimuthal component of the surface coordinate system, in the direction of the magnetic field. From Eq. (\ref{HamMagExtInt}), it can be seen that $n$ has to be even in order to minimize the energy, as well as in the radial field case. 

With these considerations, the simplified expression
\begin{eqnarray}\label{ThetaEqMagExtSimp}
\partial_{\zeta}^2\Theta=\frac{\sin2\Theta}{2}-\frac{\rho^2}{2}
{B'(\rho)}\cos\Theta\,
\end{eqnarray}
is obtained. Again, in order to get analytical solutions we must consider an AMF of the form $B'(\rho)=B'_0/\rho^2$. Thus, Eq. (\ref{ThetaRadSimp}) is recovered and the below discussions refer to the Eqs. (\ref{ThetaRadSimp}) and (\ref{ThetaEqMagExtSimp}).

\section{Results}\label{Results}

\begin{figure}
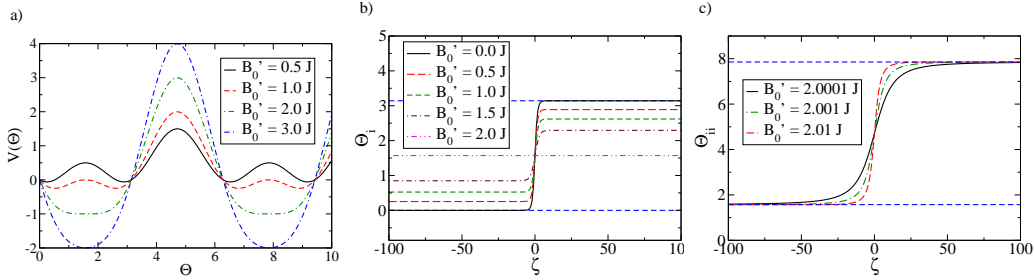

\includegraphics[scale=0.18]{potential.eps}\hspace{0.2cm}\includegraphics[scale=0.18]{weak.eps}\includegraphics[scale=0.18]{skyrmion.eps}\caption{[Color online]  Figure (a) shows the behavior of the potential given by Eq. (\ref{PotAziField}) for different values of $B_0$. For $B_0<2$, There are two minima connected by a small energy barrier. Note that the energy barrier connecting the minima $\Theta_0=\arcsin(B'_0/2)$ and $-(\pi+\Theta_0)$ is greater than the one connecting $\Theta_0$ and $\pi-\Theta_0$. Figure (b) shows the skyrmion-like profile connecting these minima. Figure (c) shows the $2\pi$-skyrmion solution given by Eq. (\ref{ThetaMagAziSol}) for different values of the magnetic field. When the magnetic field strength increases, the skyrmion width diminishes. Dotted lines represent the two vacua, $\pi/2$ and $5\pi/2$, connected by the skyrmion and are guide for the eyes.}\label{PotGraf}
\end{figure}
We start by considering the total Hamiltonian of the system given by $H_\text{T}=H+H_{\text{int}}$, and rewrite the energy density as $\mathcal{H}=(\partial_\zeta\Theta)^2+V(\Theta)$, where
\begin{eqnarray}\label{PotAziField}
V(\Theta)=\sin^2\Theta-B'_{0}\sin\Theta\,,
\end{eqnarray}
with
\begin{eqnarray}\label{DerPotAziField}
V'(\Theta)=2\sin\Theta\cos\Theta-B'_0\cos\Theta\,,
\end{eqnarray}
where $V'(\Theta)$ denotes the derivative of the potential. To obtain some insight into the solutions, we consider separately  two situations, weak magnetic field ($B'_{0}<2$), and strong magnetic field ($B'_{0}>2$). When a weak magnetic field is considered there are two equivalent minima, $\Theta_0=\sin^{-1}(B'_{0}/2)$ and $\Theta_1=\pi-\Theta_0$, as  shown in Fig. \ref{PotGraf}.(a). When we consider a weak magnetic field, Eq. (\ref{ThetaEqMagExtSimp}) describes an equivalent system as described by Eq. (2.6) in Ref. \cite{Leung-PRB-1983}. In the later, the authors considered a relatively weak field and obtain a solution that can well represent, in our case, the solution of Eq. (\ref{ThetaRadSimp}).

Therefore, our solution can be represented by the following profile obtained from  \cite{Leung-PRB-1983} in the context of the unidimensional HM
\begin{eqnarray}\label{ThetaAziField}
\Theta_{\text{i}}=2\tan^{-1}\left[\left(\frac{2-B'_{0}}{2+B'_{0}}\right)^{1/2}\tanh\left(\frac{\lambda\zeta}{2}\right)\right]+\frac{\pi}{2}\,,
\end{eqnarray}
where $\lambda^{-1}=2/(4-B'^2_{0})^{1/2}$ is the width of the excitation. The energy and charge associated to this solution on an infinity surface are, respectively
\begin{eqnarray}
E_{(\text{i})}=4\pi J\left\{\lambda+\frac{B'_{0}}{2}\left[\sin^{-1}\left(\frac{B'_0}{2}\right)-\frac{\pi}{2}\right]\right\}\,,
\end{eqnarray}
and
\begin{eqnarray}
Q_{\text{WMF(i)}}=\frac{1}{4\pi}\int\sin\Theta d\Theta d\Phi=\sin\left[2\arctan\left(\frac{2-B'_0}{2+B'_0}\right)\right]\,\leq1.
\end{eqnarray}

Eq. (\ref{ThetaAziField}) consists in a spin vector field variation connecting the two minima. However, due to its fractional charge, its stability can not be ensured using arguments from the homotopy theory. From Fig. \ref{PotGraf}.(a), one can note that the energy barrier connecting $\Theta_0$ and $-(\pi+\Theta_0)$ is greater than the one connecting $\Theta_0$ with $\Theta_1$. Therefore, the solution which connects $\Theta_0$ and $-(\pi+\Theta_0)$ is not energetically favorable and does not appear into the studied system. Then, a weak azimuthal or radial magnetic field acting on the considered system shifts the minima associated to the isotropic HM, joining them, but does not ensure the appearance of stable integer skyrmions, which are possible even when $B'_{0}\rightarrow0$. In this case,  the sine-Gordon system is recovered and the minima to the potential $V(\Theta)$ are evaluated as $\Theta_{0}=0$ and $\Theta_1=\pi$, and $Q_{(\text{i})}\rightarrow1$. On the other hand, 
 for $B'_{0}\rightarrow2$, the skyrmion-like solution width diverges and $Q_{\text{MF(i)}}\rightarrow0$, that is, the spin vector field variation is very small along the surface and $\Theta\rightarrow\pi/2$ (See Fig. \ref{PotGraf}.(b)). Thus, the presence of a weak magnetic field destabilizes the $\pi$-skyrmion appearing in spin systems described by the isotropic Heisenberg Hamiltonian by shifting the minima associated to this model. When $B'_0=2$, the minima degenerate in $\Theta\rightarrow\pi/2$ and a vortex state pointing perpendicular or parallel to the surface is obtained under the presence of a RMF or AMF, respectively.

On the other hand, by considering the case in which a strong magnetic field is acting into the system, the minima of the potential (\ref{PotAziField}) can not be given by the expression 
$\Theta_0=\sin^{-1}(B'_0/2)$. Thus, starting from Eq. (\ref{DerPotAziField}), we have that  $\Theta_{\text{min}}=(4n+1)\pi/2$, with $n$ integer, minimizes the potential and the interaction energy given by Eq. (\ref{HamMagExtInt}). In this case the solution to Eq. (\ref{ThetaEqMagExtSimp}) is presented in Ref.\cite{Leung-PRB-1983}
\begin{eqnarray}\label{ThetaMagAziSol}
\Theta_{\text{ii}}=2\sin^{-1}\sqrt{\frac{1-\tanh^2\left(\frac{\zeta}{\tau}\right)}{1+\tau^{2}\tanh^2\left(\frac{\zeta}{\tau}\right)}}+\frac{\pi}{2}\,,
\end{eqnarray}
where $\tau=[2/(B'_0-2)]^{1/2}$. The energy associated to this excitation is given by
\begin{eqnarray}
E_{(\text{ii})}=8\pi J\left(\tau^{-1}+\frac{B'_{0}}{2}\tan^{-1}\tau\right)\,.
\end{eqnarray}

Eq. (\ref{ThetaMagAziSol}) represent a $2\pi$-skyrmion with a topological charge $Q=2$ connecting the two minima $\pi/2$ and $5\pi/2$(See Fig. \ref{PotGraf}.(c)). The width of the skyrmion is given by $\lambda^{-1}=\tau$. Thus, the magnetic field induces a geometrical frustration into the system, changing the skyrmion width in such way that as $B'_0$ grows, the skyrmion width shrinks and confines the excitation to progressively smaller regions of the surface.  It can be also observed that the energy of this excitation increases with the magnetic field strength due to the skyrmion confinement. For $B\rightarrow2$, the solutions (\ref{ThetaMagAziSol}) and (\ref{ThetaAziField}) join smoothly, going both to $\Theta=\pi/2$. 

In summary, the presence of a RMF or AMF varying with $1/\rho^2$ destabilizes the $\pi$-skyrmion connecting the minima of the sine-Gordon system. Weak magnetic fields are not enough to generate a new class of skyrmions. However, when the magnetic field increases, $2\pi$-skyrmions can appear on the surface. In this case the increase of the magnetic field confine the skyrmion in  small regions of the surface, decreasing its width. Furthermore, the behavior of the system as a function of $B_0'$ is the following, the ground state is doubly degenerate for weak fields ($B_0'<2$), becomes unique at $B_0'=2$, and is again degenerate for $B_0'>2$. The connection between the minima represents a stable excitations only in the strong field case, once the spin sphere is completely mapped twice.

\section{Conclusions and prospects}\label{conclusions}

We studied spin systems on curved surfaces with rotational symmetry in the presence of radial and azimuthal magnetic fields and described by a Heisenberg Hamiltonian. The Euler Lagrange equations coming from the model  can be analytically solved only if the magnetic field varies with the curvature of the surface in the form $B'(\rho)\propto1/\rho^2$. Weak and strong magnetic fields leads to different classes of solutions to the proposed model. In the case of weak magnetic fields, skyrmionic-like profiles are predicted, connecting the two minima associated to the potential. However, these solutions have not integer skyrmion charges and arguments coming from homotopy theory can not be used to ensure their stabilities. Strong magnetic fields yield $2\pi$-skyrmion-like excitations, with an integer topological number. The width of the skyrmions decreases when the magnetic field increases. The phase of the skyrmion depends on the magnetic field direction in such way that the azimuthal component of the spins points along the field direction. Therefore, by manipulating both the magnitude and the direction of the applied magnetic field, the magnetic system can support various types of skyrmions.

Despite skyrmions appear only as an excited state, the presented results are important to give some insights in the study of helimagnets, in which there are chiral interactions, known as Dzyaloshinskii-Moriya interactions, which lead to the appearance of skyrmions even near room-temperature \cite{Nature-2010}. Furthermore, the coupling between the magnetic field and the curvature is an interesting result, opening new issues on the control of the properties of condensed matter physics by magnetic and electric fields.

\section*{acknowledgements}
V.L.C.S. thanks the Brazilian agency CNPq (Grant No. 229053/2013-0), for financial support. J.M.F. thanks the support of FAPEMIG. D.A. acknowledges the support of FONDECYT under projects 1120356, the Milennium Science nucleus ``Basic and Applied Magnetism'' P10-161-F from MINECON, and Financiamento Basal para Centros Cient\'ificos e tecnol\'ogicos de Escelencia, under project FB 0807. We also acknowledge AFOSR Grant. No. FA9550-11-1-0347. R.G.E. thanks Conicyt Pai/Concurso Nacional de Apoyo al Retorno de Investigadores/as desde el Extranjero Folio 821320024.

\end{document}